\documentclass[%
 reprint,
superscriptaddress,
 amsmath,amssymb,
 aps,
pra,
]{revtex4-1}

\usepackage{graphicx}% Include figure files
\usepackage{subfigure}
\usepackage{overpic}
\usepackage{dcolumn}% Align table columns on decimal point
\usepackage{bm}% bold math
\usepackage{color}% textcolor
\usepackage{soul}% st
\usepackage{hyperref}% add hypertext capabilities
\begin{document}

\preprint{APS/123-QED}

\title{Non-equilibrium Dynamics of Fermi Polarons Driven by Time-dependent Interaction}% Force line breaks with \\
%\thanks{A footnote to the article title}%

\author{Fan Yang}
\affiliation{School of Physics, Renmin University of China, Beijing, 100872, P. R. China}
\affiliation{Key Laboratory of Quantum State Construction and Manipulation (Ministry of Education), Renmin University of China, Beijing, 100872, China}
 %\altaffiliation[Also at ]{Physics Department, XYZ University.}%Lines break automatically or can be forced with \\
\author{Ran Qi}%
 \email{qiran@ruc.edu.cn}
\affiliation{School of Physics, Renmin University of China, Beijing, 100872, P. R. China}
\affiliation{Key Laboratory of Quantum State Construction and Manipulation (Ministry of Education), Renmin University of China, Beijing, 100872, China}%

\date{\today}% It is always \today, today,
             %  but any date may be explicitly specified

\begin{abstract}
We systematically studied the dynamics of Fermi polarons driven by time-dependent scattering length using a time-dependent variational method. Starting from the non-interacting initial state, we calculated the evolution behavior of the contact, energy, and quasiparticle residue of this system as the scattering length $a_s(t)$ increases from zero. In the short-time evolution, we obtained analytical results, verifying that when $a_s(t)$ grows as $\sqrt{t}$, the contact $C(t)$ and energy $E(t)$ exhibit the maximum growth rate. Furthermore, we numerically solved the long-time evolution, when $a_s(t)$ is diven with a strength $\beta$ and a power $\alpha$ as $a_s(t) = {\sqrt{2}\beta}/{k_F} {(\epsilon_Ft)}^{\alpha}$. For large driving strength, due to the interference between the polaron states and the non-interacting states, $C(t)$ and $Z(t)$ exhibit oscillatory behavior; For small $\beta$, the oscillatory behavior disappears due to the decay of the repulsive polaron into the continuum, and $C(t)$ and $Z(t)$ gradually relax to zero. The energy always saturates over long times for different driving strength. Additionally, our method is applicable to any time-dependent form of the scattering length.
%\begin{description}
%\item[Usage]
%Secondary publications and information retrieval purposes.
%\item[Structure]
%You may use the \texttt{description} environment to structure your abstract;
%use the optional argument of the \verb+\item+ command to give the category of each item.
%\end{description}
\end{abstract}

%\keywords{Suggested keywords}%Use showkeys class option if keyword
                              %display desired
\maketitle

%\tableofcontents

\section{\label{sec:level1}Introduction}
The concept of polaron quasiparticle is first introduced by Landau and Pekar to describe the an electron dressed by the lattice phonons when moving through lattice ions. \cite{LP1, LP2}
Ultracold atomic gases have extended this idea, offering an analogous platform to explore Fermi polarons -- quasiparticles that emerge when an impurity atom interacts with a surrounding Fermi sea, forming a correlated cloud of particle-hole excitations \cite{exp1, exp2,exp3,exp4,exp5,exp6,exp7,exp8,exp9}.
Fermi polaron systems are important for understanding the interplay between impurity and many-body physics. Its phase diagram as well as quasi-particle properties such as energy, effective mass, and spectral features, have been extensively investigated theoretically by different approaches \cite{theo1,theo2,theo3,theo4,theo5,theo6,theo7,theo8,theo9,theo10,theo11,theo12,
theo13,theo14,theo15}.

Ultracold gases provide an unprecedented level of control over these systems, making them ideal for probing beyond-equilibrium physics. The ability to precisely tune interactions via Feshbach resonances enables the exploration on a wide range of dynamical regimes, from adiabatic to quench processes \cite{quench1,quench2,quench3}, and dynamics under periodically modulating interactions \cite{period1,period2,period3,period4}.
Compared to their solid-state counterparts, ultracold systems operate on far longer timescales for many-body response due to the lower Fermi energy $\epsilon_F$, allowing detailed studies of dynamical processes. This has enabled experimental exploration of phenomena like quasiparticle formation and excitation spectra after sudden interaction quenches \cite{quench3}. Theoretical approaches such as the functional determinant approach (FDA) \cite{FDA} and truncated basis methods (TBM) \cite{TBM1,TBM2} have been developed to model these dynamics.

While much progress has been made in understanding quench dynamics, the broader nonequilibrium behavior of Fermi polarons, especially under continuous time-dependent interaction ramping, remains an open study. Recent theoretical work \cite{Qi} has revealed intriguing cases, such as universal maximum energy growth rates of a quantum gas under specific interaction ramps, i.e. the two-body scattering length $a_s(t)$ increases as $\sqrt{t}$, when the short-time Schr\"{o}dinger equation remains scale invariant.

In this paper, we use a dynamic variational approach to study the many-body dynamics of the Fermi polaron system driven by time-dependent interaction. The $s$-wave scattering length $a_s(t)$ between the impurity and the host fermion is tuned to be increased from zero with an arbitrary time dependence. This gives a time-dependent low-energy scattering amplitude $f_k(t)$ and thus a $T$-matrix $T_E(t)$ as
\begin{equation}
  T_E(t) = -\frac{2\pi\hbar^2}{m_r}f_k(t) = \frac{2\pi\hbar^2}{m_r}\frac{1}{1/a_s(t) + ik},\label{eq.time-T}
\end{equation}
where $m_r$ is the reduced mass between the impurity and the host fermion, $k=\sqrt{2m_rE/\hbar^2}$, and $E$ is the scattering energy.

We adopt a time-dependent Chevy's ansatz and confine the evolution of the wave functions to a subspace with up to one particle-hole excitation. The one particle-hole approximation remains valid for long times as long as the gases are sufficiently dilute.
We focus on the cases that $a_s(t)$ is diven with a strength $\beta$ and a power $\alpha$ as $a_s(t) = {\sqrt{2}\beta}/{k_F} {(\epsilon_Ft)}^{\alpha}$, where $\epsilon_F$ is the Fermi energy of the background Fermi sea. Specifically, we calculate the dynamical behaviors of the contact, energy, and quasiparticle residue of the Fermi polaron. We analytically check the universal behaviors in short times. For long times, we observe two distinct dynamic behaviors as the driving strength increases. For large $\beta$, the contact and quasiparticle residue exhibit damped oscillation around a finite relaxation value, while for small $\beta$, the contact and quasiparticle residue decay to zero at long time.

The remainder of this paper is organized as follows. In Sec. \ref{sec:level2} we introduce the our system and the dynamical variational approach. In Sec. \ref{sec:level3}, we derivate the time-dependent expression of the contact and other quasiparticle quantities with the variational parameters. In Sec. \ref{sec:level4}, we show the obtained analytical results at short time and the numerical results at long time. A summary of this work is given in Sec. In Sec. \ref{sec:level5}.

\section{Model and Dynamical Variational Approach\label{sec:level2}}

\subsection{Hamiltonian}
We study the dynamics of a single impurity interacting with a Fermi sea via a time-dependent two-body interaction.  Then the Hamiltonian of this many-body system can be writen as
\begin{equation}
  \hat H = \sum_{\mathbf{k}\sigma} \epsilon_{\mathbf{k},\sigma} c_{\mathbf{k},\sigma}^\dagger c_{\mathbf{k},\sigma} + \frac{g(t)}{V}\sideset{}{_{\mathbf{k}\mathbf{k}'\mathbf{q}}^{\prime}}\sum c_{\mathbf{k}-\mathbf{q},\uparrow}^{\dagger}c_{\mathbf{k}'+q,\downarrow}^{\dagger}c_{\mathbf{k}',\downarrow}c_{\mathbf{k},\uparrow}
\end{equation}
Here $\hat{c}_{\mathbf{k},\sigma}^{\dagger}$ ($\hat{c}_{\mathbf{k},\sigma}$) is the creation (annihilation) operator for particles with momentum $\mathbf{k}$ and the mass $m_{\sigma}$, where the spin $\sigma = \downarrow$ (or $\uparrow$) labels the impurity (or host fermions);
$\epsilon_{\mathbf{k},\downarrow} =k^2/(2m_{\downarrow})$ and $\epsilon_{\mathbf{k},\uparrow} = k^2/(2m_{\uparrow})-\epsilon_F$ donates the non-interacting dispersions, where $\epsilon_F=k_F^2/(2m_{\uparrow})$ with $k_F$ relating to the density of the fermions $n_{\uparrow}$ as $k_F^3=6\pi^2n_{\uparrow}$.
According to Eq. (\ref{eq.time-T}), one can write down the following time-dependent renormalization relation between $g(t)$ and the time-dependent scattering length $a_s(t)$ as
\begin{equation}
  \frac{1}{g(t)}=\frac{m_r}{2\pi\hbar^2 a_s(t)}-\frac{1}{V}\sum_{\mathbf{k}}\frac{1}{\hbar^2k^2/(2m_r)},\label{eq.g-renormal}
\end{equation}
and $m_r=m_{\uparrow}m_{\downarrow}/(m_{\uparrow}+m_{\downarrow})$. In this work, we consider the equal-mass case, such as $m_{\downarrow}=m_{\uparrow}= 2m_r= m$, and set $\hbar\equiv m\equiv 1$ in the rest of this paper.

\subsection{Dynamical Variational Approach}
To study the many-body dynamics, we adopt a time-dependent variational approach\cite{dynvar1,dynvar2,dynvar3,dynvar4}, which confines the evolution of wave-functions to a subspace of the many-body Hamiltonian. The wave function is chosen as a time-dependent Chevy's ansatz\cite{theo1} up to the one particle-hole order, and is given by
\begin{equation}
  |\Psi(t)\rangle = \left(\varphi_0(t) c_{\mathbf{p}=0,\downarrow}^{\dagger} + \sideset{}{_{\mathbf{kq}}^{\prime}}\sum\varphi_{\mathbf{kq}}(t)c_{\mathbf{q}-\mathbf{k},\downarrow}^{\dagger}c_{\mathbf{k},\uparrow}^{\dagger}c_{\mathbf{q},\uparrow}\right)|FS\rangle,\label{eq.psi}
\end{equation}
where $\varphi_0$ and $\varphi_{\mathbf{kq}}$ are variational parameters, and $\sum^{\prime}$ refers to as $\sum_{|\mathbf{k}|>k_F}$ and $\sum_{|\mathbf{q}|<k_F}$. These variational parameters determine the occupations of the impurity on zero-momentum and finite momentum $\mathbf{k}$ via $n_{0,\downarrow}(t)=|\varphi_0(t)|^2$ and $n_{\mathbf{k},\downarrow}(t)=\sum_{|\mathbf{q}|<k_F,|\mathbf{q}-\mathbf{k}|>k_F}|\varphi_{\mathbf{q}-\mathbf{k},\mathbf{q}}|^2$.
The evolution of the variational parameters can be obtained from the Euler-Lagrange equation for the Lagrangian $\mathcal{L}=\frac{i}{2}[\langle\Psi(t) \mid \dot{\Psi}(t)\rangle-\langle\dot{\Psi}(t) \mid \Psi(t)\rangle]-\langle\Psi(t)|\hat{H}(t)| \Psi(t)\rangle$ as follows
\begin{subequations}
	\begin{align}
	    i\dot{\varphi}_{0}(t) &= \frac{1}{V}\sideset{}{_{\mathbf{q}}^{\prime}}\sum\chi_q(t),\\
	    i\dot{\varphi}_{\mathbf{kq}}(t) &= E_{\mathbf{kq}}\varphi_{\mathbf{kq}}(t) + \frac{1}{V}\chi_q(t),
	\end{align}\label{eq.phi-derivative}
\end{subequations}
where $E_{\mathbf{kq}}= \epsilon_{\mathbf{k},\downarrow} + \epsilon_{\mathbf{q}-\mathbf{k},\uparrow} -\epsilon_{\mathbf{q}-\mathbf{k},\uparrow}$ relates to the kinetic energy increase of a particle-hole excitation, and the auxiliary function $\chi_q(t)$ is defined as $\chi_q(t)=g[\varphi_0(t)+\sum_{\mathbf{k}}'\varphi_{kq})$. It is easy to check that the particle number for both impurity and Fermi sea is conserved. We take non-interacting state as the initial state, such as $\varphi_0(0)=1$ and $\varphi_{\mathbf{kq}}(0)=0$. The integral of Eq.~(\ref{eq.phi-derivative}) combimed with the renormalization relation Eq.~(\ref{eq.g-renormal}) lead to the the integral equation of $\chi_q(t)$ as
%\begin{subequations}
%  \begin{align}
%    \varphi_0(t) &= 1 - i\frac{1}{V}\sideset{}{_{\mathbf{q}^{\prime}}^{\prime}}\sum\int_0^t\mathrm{d}\tau\chi_{q'}(\tau),\label{eq.phi0}\\
%    \varphi_{\mathbf{kq}}(t) &= - i\frac{1}{V}\int_0^t\mathrm{d}\tau\chi_q(\tau)e^{-iE_{\mathbf{kq}}(t-\tau)},\label{eq.phikq}
%  \end{align}
%\end{subequations}
\begin{align}
  \left[\frac{1}{4\pi a_s(t)}+\hat{L}\right]\chi_q(t) = 1- i \frac{1}{V}\sum_{q'}\int_0^t\mathrm{d}\tau\chi_{q'}(\tau),\label{eq.chi_q}
\end{align}
where $\hat{L}$ is a linear operator acting on $\chi_{q}(t)$ as
\begin{equation}
  \hat{L}\chi_q(t) = \lim_{\epsilon\to 0}i\left[\int_0^{t-\epsilon}\!\mathbf{d}\tau f_q(t\!-\!\tau)\chi_q(\tau) + \frac{\sqrt{i}}{4 \pi^{3/2}}\frac{\chi_q(t)}{\sqrt{\epsilon}}\right],
\end{equation}
and the kernel $f_q(t)$ is given by
\begin{equation}
  f_q(t) = \frac{1}{V} \sideset{}{_{\mathbf{k}}^{\prime}}\sum\exp(-iE_{\mathbf{kq}}t).
\end{equation}

\section{Derivation of the physical quantities\label{sec:level3}}

The contact $C(t)$ can be obtained from the instantaneously momentum distribution \cite{appA} as.
\begin{equation}
  C(t) = n_{\downarrow}\frac{1}{V} \sideset{}{_{\mathbf{q}}^{\prime}}\sum|\chi_q(t)|^2,\label{eq.contact}
\end{equation}
where $n_{\downarrow}=1/V$ is the particle density of the impurity. The relation between contact and $\chi_q$ is similar with that in the equilibrium case. Then the dynamics of energy $E$ can be obtained by Tan's sweep theorem \cite{Tansweep} as
\begin{equation}
  \frac{1}{V}\frac{d}{dt}E(t) = \frac{C(t)}{4\pi a^2(t)}\frac{da}{dt}.\label{eq.energy}
\end{equation}
Another important quantity of Fermi polaron problem is the quasi-particle residue $Z$. In this variational method, $Z(t)$ is defined by the overlap with the noninteracting impurity state as
\begin{equation}
  Z(t) = |\varphi_0(t)|^2,\label{eq.residue}
\end{equation}
and $\varphi_0(t)$ is related to $\chi_q(t)$ as Eq. (\ref{eq.phi-derivative}a). 
In a Ramsey interference expriment, $\varphi_0(t)$ can be measured by the Ramsey contrast, which has been used as a probe of many-body dynamics \cite{Loschmidt}. %The definition of $S(t)$ is analogous to that of the Loschmidt echo, given by
%\begin{equation}
%  S(t) = \langle\Psi(0)|U_0^{-1}(t)U(t)|\Psi(0)\rangle = \varphi_0(t),\label{eq.contrast}
%\end{equation}
%where $U_0(t)$ donates the free evolution without interaction, $U(t)=\exp[-i\mathcal{T}\int_0^t\hat{H}(\tau)\mathrm{d}\tau]$ is the evolution operator under $H(t)$, and we set the energy of non-interacting Fermi sea as the energy zero point.

In the section below, we show $\chi_q(t)$ can be obtained analytically at short time and numerically at long time by solving Eq.~(\ref{eq.chi_q}). Substituding the solution into Eqs. (\ref{eq.contact}) and (\ref{eq.phi-derivative}a), we can obtain the contact $C(t)$ and $\varphi_0(t)$, and subsequently, the dynamics of energy $E(t)$ and quasi-particle residue $Z(t)$ via Eqs. (\ref{eq.energy}-\ref{eq.residue}).

\section{Results\label{sec:level4}}
\subsection{Short-time Dynamics}
We consider that $a_s(t)$ grows from zero to a positive value, and without loss of generality, is a power-law function of $t$ as
\begin{equation}
  a_s(t) = \frac{\sqrt{2}\beta}{k_F} {(\epsilon_Ft)}^{\alpha}, \quad \alpha>0.\label{eq.as-t}
\end{equation}
At short time ($t\ll t_F =1/\epsilon_F$), the operator has an important property that
\begin{align}
  \hat{L}x^{\gamma} =& i\sum_{n}b_{n}B(\gamma,n)x^{\gamma+n+1},\notag\\
  &\ \quad n=-3/2,-1/2,0,1/2,1,\dots.\label{eq.L}
\end{align}
where the coefficients $b_{n}$ are that of the Taylor expansion of $f(t)$ at $t\to 0$, and $B(x,y)=\Gamma(x+1)\Gamma(y+1)/\Gamma(x+y+2)$ with $\Gamma(x)$ being the gamma function. Then, Eq.(\ref{eq.chi_q}) can be solved analytically by assuming the asymptotic solution has the form of $\chi_q(t) = \sum_{\lambda} c_{\lambda}t^{\lambda}$. The leading term of $\chi_q(t)$ depends on $\alpha$ and is determined by whether the $1/a_s(t)$ term or $\hat{L}$ operator dominates Eq.~(\ref{eq.chi_q}). 

Case I: $\alpha>1/2$. In this case, $1/a_s(t)$ term dominates, the short time dynamics is consistent with the adiabatic regime.To the leading order of $t$, contact and energy only depend on the instantaneous scattering length at time $t$ as
\begin{subequations}
   \begin{align}
     C(t) &\approx 16\pi^2n_{\uparrow}n_{\downarrow}a_s^2(t).\\
     E(t) &\approx 4\pi n_{\uparrow}a_s(t),
   \end{align}
\end{subequations}

Case II:  $0<\alpha<1/2$. The $\hat{L}$ operator dominates, and the leading order term of $C(t)$ and $E(t)$ are given by
\begin{subequations}
   \begin{align}
      C(t) &= 64\pi n_{\uparrow}n_{\downarrow}t,\\
      E(t) &= \frac{16\sqrt{2}\alpha}{3\pi^2\beta(1-\alpha)}\epsilon_F{(\epsilon_Ft)}^{1-\alpha}
   \end{align}
\end{subequations}
The short time for this case is the same as a quench process, since the growth of contact and energy is independent on parameters $E_F$ or $\beta$, and contact always grows linearly in time with a constant rate no matter how fast $a_s(t)$ increases at initial.

Case III:   $\alpha=1/2$. For this critical condition, $a_s(t) = \beta\sqrt{t}$ is independent on $k_F$, and the $\hat{L}$ term and $1/(4\pi a_s)$ are equally important,
which gives the leading order of $C(t)$ and $E(t)$ as
\begin{subequations}
   \begin{align}
      C(t) &= n_{\uparrow}n_{\downarrow}\left|c_{1/2}(\beta)\right|^2t,\\
      E(t) &= n_{\uparrow}\frac{|c_{1/2}(\beta)|^2}{4\pi\beta}\sqrt{t},
   \end{align}
\end{subequations}
where
\begin{equation}
    c_{1/2}(\beta) = 4\pi\left(\frac{1}{\beta}+\frac{B(1/2,-3/2)}{\sqrt{4\pi i}}\right)^{-1}.
\end{equation}

\begin{figure}[htbp]
  \centering
  \includegraphics[width=0.99\linewidth]{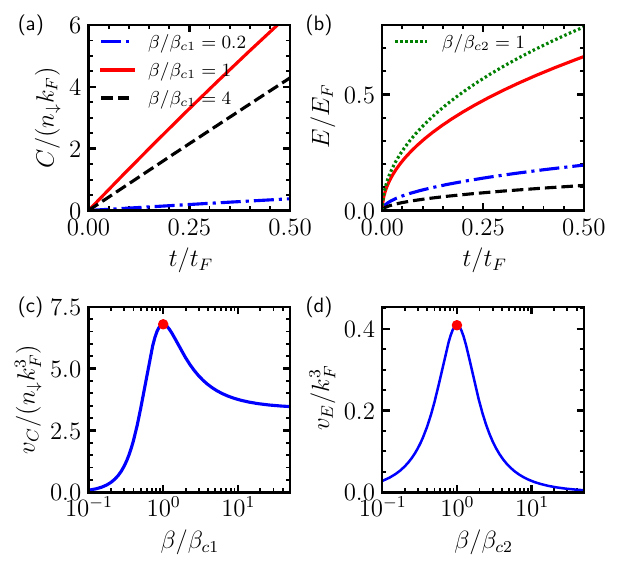}
  \caption{$\textbf{(a-b)}$: The short time evolution of contact and energy for $\alpha=1/2$ with different $\beta$. $\textbf{(c-d)}$: The initial growth rate for contact and energy for $\alpha=1/2$ as a function of $\beta$.}\label{fig.CE-short}
\end{figure}

With definition of the initial growth rate for contact and energy as $v_C=\lim_{t\to 0}\mathrm{d}C(t)/\mathrm{d}t$ and $v_E=\lim_{t\to 0}\mathrm{d}E(t)/\mathrm{d}\sqrt{t}$, the contact growth and energy growth at the short time are fastest in the case $\alpha=1/2$. As shown in Fig.~\ref{fig.CE-short}, $v_C$ reaches its maximum at $\beta_{c1}=2\sqrt{2/\pi}$, while $v_E$ at $\beta_{c2}=2/\sqrt{\pi}$.
The leading-order results above obtained by the variational approach in this Fermi polaron system are consistent with those from a universal calculation reported in Ref. \cite{Qi}.

\subsection{Long-time Dynamics}
We solve Eq. (\ref{eq.chi_q}) numerically for $\chi_q$ at long time and calculate long-time dynamics. Interestingly, for both the contact $C(t)$ and $Z(t)$, we obtain two distinct dynamical behaviors at different $\beta$.

\begin{figure}[htbp]
  \centering
  \includegraphics[width=0.99\linewidth]{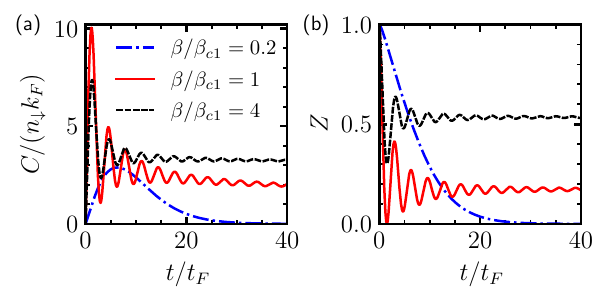}
  \caption{The long time evolution of $\textbf{(a)}$ contact and $\textbf{(b)}$ quasi-particle residue for $\alpha=1/2$ with different $\beta$.}\label{fig.CZ-long}
\end{figure}
Taking the case of $\alpha=1/2$ as example,  Fig. \ref{fig.CZ-long}(a) shows the numerical results of $C(t)$. At larger $\beta$, $C(t)$ exhibits damped oscillation around a relaxation value, which becomes larger as $\beta$ increased. The oscillation reflects the well-defined polaron states, repulsive and attractive polaron, which both exist at $a_s>0$ and have a finite overlap with the non-interacting initial state. As $\beta$ decreases, the long-time behavior crossovers to continuous decay towards 0. These results may be attributed to the finite lifetime of the repulsive polaron. For small $\beta$, at the early scenario, the system adiabatically evolves along the repulsive polaron branch, and the attractive polaron hardly involves in this process. However, due to the instability of the repulsive polaron, the system may decay to the dressed dimer continuum resulting in two consequences. On the one hand, the oscillation can be smoothed out due to the large population on the continuum of dressed dimers. On the other hand, $Z(t)$ should reduce to zero at long time, as the dimer states have zero overlap with the non-interacting state. 

\begin{figure}[htbp]
  \centering
  \includegraphics[width=0.90\linewidth]{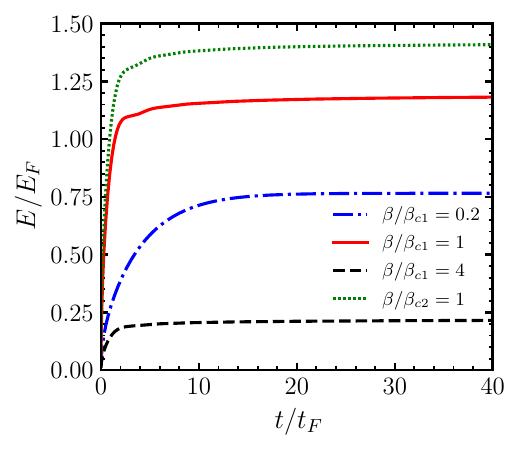}
  \caption{The long time evolution of the energy for $\alpha=1/2$ with different $\beta$.}\label{fig.e-long}
\end{figure}

Fig.~\ref{fig.CZ-long}(b) shows the long-time dynamics of quasi-particle residue $Z(t)$. Indeed,  $Z(t)$ also exhibit damped oscillation for large values of $\beta$, while for small $\beta$, it decays to zero over time. Finally, we calculate $E(t)$ through the Tan's sweep theorem. As shown in Fig.\ref{fig.e-long}, $E(t)$ increases monotonically, but saturates over a long time. This can be seen from Eq. (\ref{eq.energy}), since at long time, $C(t)$ relaxes to a finite value and $\left(da_s/dt\right)/a_s^{2}$ decays faster than $1/t$ for $a_s(t)\propto t^{\alpha}$ with any positive $\alpha$.

\begin{figure}[htbp]
  \centering
  \includegraphics[width=0.99\linewidth]{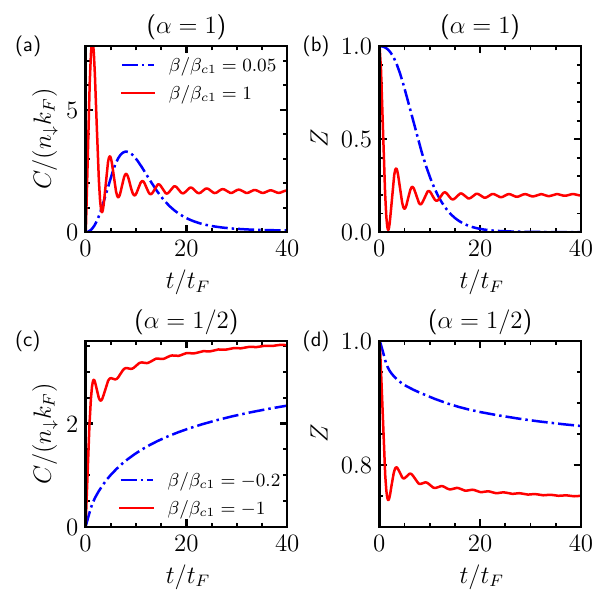}
  \caption{$\textbf{(a-b)}$: The long time evolution of contact and residue for $\alpha=1$ with $\beta=0.05$ and 1. $\textbf{(c-d)}$: The long time evolution of contact and residue for $\alpha=1/2$ with $\beta=-0.2$ and $-1$.}\label{fig.linear}
\end{figure}

We also calculate the cases for other positive values of $\alpha$, e.g. for $\alpha=1$ as shown in Fig. \ref{fig.linear}. The results indicate that, as $\beta$ decreases, the transition in long-time dynamic behavior from damped oscillations to a continuous decay to zero is not unique to the case of $\alpha=1/2$.
Fig. show the results for small but negative $\beta$, the relaxation value of contact and residue are on longer vanish since the attractive polarons are stable. 

\section{Conclusion\label{sec:level5}}
We use a time-dependent variational approach to explore the non-equilibrium dynamics of a Fermi polaron system driven by time-dependent interactions. The short-time dynamics are derived analytically, with the maximum growth rates of contact and energy reproducing the universal results reported in Ref\cite{Qi}. Long-time dynamics are numerically investigated under various scenarios. For strong driving, the quasiparticle residue survives at long time, and the contact relaxes to finite value. For weak driving, short-time adiabaticity and the instability of repulsive polaron lead to the decay of both the contact and residue to zero at long times.

\begin{acknowledgments}
This work was supported by the National Key Research and Development Program of China (Grant  No. 2022YFA1405301 (R.Q.) and No. 2018YFA0306502 (R.Q.)), 
the  National Natural Science Foundation of China (Grant No. 12022405 (R.Q.) and No. 11774426 (R.Q.)), and
the Beijing Natural Science Foundation (Grant No. Z180013 (R.Q.)).
\end{acknowledgments}

\appendix

\section{Derivation of Eq.~(\ref{eq.contact})\label{app1}}

In this appendix, we derive the evolution of contact $\mathcal{C}(t)$ from the variation parameters. We focus on the timescale dominated by short-range two-body interactions, the instantaneous contact is given by definition as
\begin{equation}
  \mathcal{C}(t) \equiv \lim_{k\to \infty}k^4 n_{\mathbf{k},\downarrow}(t),\label{eq.ct}
\end{equation}
where $n_{\mathbf{k},\downarrow}(t)$ is the particle number of the impurity with the momentum $\mathbf{k}$ at $t$. For $|\mathbf{k}|>2k_F$, we have $n_{|\mathbf{k}|>2k_F,\downarrow}(t)=\sum_{\mathbf{q}}^{\prime}|\varphi_{\mathbf{kq}}(t)|^2$.
With the non-interacting initial state, {\it e.g.} $\varphi_{\mathbf{kq}}(0)\equiv 0$, Eq.~(\ref{eq.phi-derivative}b) gives
\begin{align}
  \varphi_{\mathbf{kq}}(t) &= - i\frac{1}{V}\int_0^t\chi_q(\tau)e^{-iE_{\mathbf{kq}}(t-\tau)}\mathrm{d}\tau \label{eq.phikq-t}
\end{align}

Assuming $\chi_q(t)$ can be expanded as
\begin{equation}
  \chi_q(t) = \sum_{\lambda} b_{\lambda}t^{\lambda},\quad (\lambda>0),
\end{equation}
and this can be checked by solving (\ref{eq.chi_q}). As a result, $\varphi_{\mathbf{kq}}(t)$ can be rewrited by
\begin{align}
  \varphi_{\mathbf{kq}}(t) &= - i\frac{1}{V}\int_0^t\sum_{\lambda} b_{\lambda}\tau^{\lambda}e^{-iE_{\mathbf{kq}}(t-\tau)}\mathrm{d}\tau\notag\\
  &= - i\frac{1}{V}\sum_{\lambda}b_{\lambda}t^{\lambda+1}\int_0^1 x^{\lambda}e^{-iy(1-x)}\mathrm{d}x,
\end{align}
where $y=E_{\mathbf{kq}}t$. Then for large momentum, $y\to\infty$ as $k\to\infty$, and
\begin{eqnarray}
  \lim_{k\to \infty}\varphi_{\mathbf{kq}}(t) &\overset{y\to \infty}{\approx}& - i\frac{1}{V}\sum_{\lambda}b_{\lambda}t^{\lambda+1}\cdot\left(-\frac{i}{y}\right),\quad y=E_{\mathbf{kq}}t\notag\\
  &=& -\frac{1}{E_{\mathbf{kq}}}\frac{1}{V}\sum_{\lambda}b_{\lambda}t^{\lambda}\notag\\
  &\approx& -\frac{1}{k^2}\frac{1}{V}\chi_q(t).\label{eq.phikq-bigk}
\end{eqnarray}
which gives the particle number at $k\to\infty$ as
\begin{equation}
  \lim_{k\to\infty}n_{\mathbf{k},\downarrow}(t) = \frac{1}{k^4}\frac{1}{V^2}\sideset{}{_\mathbf{q}^{\prime}}\sum|\chi_q(t)|^2.
\end{equation}
Substituding this limit into Eq.~(\ref{eq.ct}), we obtain $\mathcal{C}(t)$ as
\begin{equation}
  \mathcal{C}(t) \equiv \lim_{k\to \infty}k^4 n_{\mathbf{k},\downarrow}(t) = n_{\downarrow}\frac{1}{V}\sideset{}{_\mathbf{q}^{\prime}}\sum|\chi_q(t)|^2.
\end{equation}
where $n_{\downarrow}=1/V$ is the density of the impurity.

\section{Ramsey procedure with the dynamical variational description}
The Ramsey interferometry expriments relies on the radio-frequency (RF) fields which can transform impurity between two hyperfine spin states $|\uparrow\rangle, |\downarrow\rangle$. The $|\uparrow\rangle$ state interacts with the host fermions via a contact otential, while the $|\downarrow\rangle$ state is noninteracting.
In this appendix, we describe the Ramsey procedures by taking the time-dependent wave-function with up to one particle-hole excitation.
\begin{equation}
  |\Psi(t)\rangle = \sum_{\sigma}\left(\varphi_{0,\sigma}(t) c_{0,\sigma}^{\dagger} + \sum_{\mathbf{kq}}'\varphi_{\mathbf{kq,\sigma}}(t)c_{q-k,\sigma}^{\dagger}c_{k}^{\dagger}c_{q}\right)|FS\rangle,
\end{equation}
where $\sigma$ labels the inner state of impurity, and $c^\dagger,c$ without $\sigma$ are creation and annihilation operators of host fermions.

In the Ramsey sequence, the impurity is initially prepared in the noninteracting down state $|\downarrow\rangle$, and after a $\pi/2$ pulse is applied, the impurity is driven into the superposition state $\frac{|\downarrow\rangle+|\uparrow\rangle}{\sqrt{2}}$. The total wave-function is approximated by
\begin{equation}
  \begin{aligned}
    |\Psi(0)\rangle &= |\Psi_1(0)\rangle + |\Psi_2(0)\rangle,\\
    |\Psi_1(0)\rangle &= \frac{1}{\sqrt{2}}c_{0,\downarrow}^{\dagger}|FS\rangle,\\
    |\Psi_2(0)\rangle &= \frac{1}{\sqrt{2}}e^{-i\phi}c_{0,\uparrow}^{\dagger}|FS\rangle,\\
  \end{aligned}
\end{equation}
which, after an evolution of t, becomes
\begin{equation}
  \begin{aligned}
    |\Psi(t)\rangle &= |\Psi_1(t)\rangle + |\Psi_2(t)\rangle,\\
    |\Psi_1(t)\rangle &= \frac{1}{\sqrt{2}}c_{0,\downarrow}^{\dagger}|FS\rangle,\\
    |\Psi_2(0)\rangle &= \frac{e^{-i\phi}}{\sqrt{2}}\left(\varphi_{0}(t) c_{0,\uparrow}^{\dagger} + \sum_{\mathbf{kq}}'\varphi_{\mathbf{kq}}(t)c_{q-k,\uparrow}^{\dagger}c_{k}^{\dagger}c_{q}\right)|FS\rangle.\\
  \end{aligned}
\end{equation}
Then, after a second $\pi/2$ pulse, we can rewrite
\begin{equation}
  c_{\mathbf{k},\downarrow}^{\dagger} \rightarrow \frac{c_{\mathbf{k},\downarrow}^{\dagger}+c_{\mathbf{k},\uparrow}^{\dagger}}{\sqrt{2}}\ ,\ c_{\mathbf{k},\uparrow}^{\dagger} \rightarrow \frac{c_{\mathbf{k},\downarrow}^{\dagger}-c_{\mathbf{k},\uparrow}^{\dagger}}{\sqrt{2}},
\end{equation}
the populations between two states of the impurity is given by
\begin{align}
  n_{\downarrow}-n_{\uparrow} &= \frac{2}{4}\left[\varphi_0(t)e^{-i\phi}+h.c.\right]=\mathrm{Re}\left[\varphi_0(t)e^{-i\phi}\right].
\end{align}
Ramsey signal $S(t)$, donated by $\varphi_0(t)$, can be measured as the populations differences,
\begin{equation}
  \frac{n_{\downarrow}-n_{\uparrow}}{n_{\downarrow}+n_{\uparrow}} = \mathrm{Re}[e^{-i\phi}S(t)].
\end{equation}

%\newpage

% The \nocite command causes all entries in a bibliography to be printed out
% whether or not they are actually referenced in the text. This is appropriate
% for the sample file to show the different styles of references, but authors
% most likely will not want to use it.
\nocite{*}

%\bibliography{ref}% Produces the bibliography via BibTeX.

\end{document}